\documentclass[aps,prl,longbibliography,twocolumn,floatfix,superscriptaddress,nofootinbib]{revtex4-2}

\usepackage{graphicx}
\usepackage{dcolumn}
\usepackage{bm}
\usepackage{latexsym,epsfig}
\usepackage{graphicx}
\usepackage{verbatim}
\usepackage{comment}
\usepackage{amsmath}
\usepackage{amsfonts}
\usepackage{amsthm}
\usepackage{amssymb}
\usepackage{mathrsfs}
\usepackage{stmaryrd}
\usepackage{color}
\usepackage{physics}
\usepackage{epstopdf}
\usepackage{grffile}
\usepackage{bbold}
\usepackage[normalem]{ulem}
\DeclareGraphicsExtensions{.eps}

\makeatletter
\newcommand{\fmarki}{$\ensuremath{\dagger}$}
\newcommand{\fmarkii}{$\ensuremath{*}$}
\newcommand{\fmarkiii}{$\ddagger$}

\def\@fnsymbol#1{{\ifcase#1\or \fmarki\or  \fmarkii\or \fmarkiii\else\@ctrerr\fi}}
\makeatother

\usepackage{hyperref}
\hypersetup{
    colorlinks=true,      
    urlcolor=blue,
    citecolor=blue,
    linkcolor=blue
}

\begin{document}

\title{Emergent criticality in the Aubry-Andr\'e model with periodic modulation}

\begin{abstract}
The Aubry-Andr\'e–Harper (AAH) model describes a system with quasiperiodic lattice modulation. In one dimension the AAH model is known to exhibit a sharp metal to insulator transition at a self-dual critical point at which all the states in the spectrum are critical or multifractal in nature. While such criticality is immediately destroyed by an additional onsite periodic modulation, we show an emergent criticality in the limit of strong periodic modulation strength under proper conditions.
The resulting strong-modulation critical phase exhibits multifractal eigenstates and singular continuous spectra, belonging to the universality class of the critical Harper model. Moreover, we reveal that additional  periodic potential of period $N$ in the quasiperiodic chain folds the spectrum into $N$ bands with quasiperiodicity being enhanced by a factor of $N$, producing $N$ numbers of Hofstadter butterflies in each band.  Our results reveal a general mechanism for engineering robust criticality and spectral replication in quasiperiodic systems.    
\end{abstract}

\author{Sitaram Maity}
\email{sitaram.maity@niser.ac.in}
\thanks{These authors contributed equally}
\author{Nilanjan Roy}
\email{nilanjanroy@niser.ac.in}
\thanks{These authors contributed equally}
\author{Tapan Mishra}
\email{mishratapan@niser.ac.in}

\affiliation{School of Physical Sciences, National Institute of Science Education and Research, Jatni,  Odisha 752050, India}
\affiliation{Homi Bhabha National Institute, Training School Complex, Anushaktinagar, Mumbai, Maharashtra 400094, India}

\date{\today}

\maketitle

\emph{Introduction.--} Quasiperiodic lattices provide an unique arena in which localization, criticality, multifractality and spectral topology arise in deterministic Hamiltonians without random disorder~\cite{jagannathan2021fibonacci}. A paradigmatic example in this context is the one dimensional Aubry--Andr\'e--Harper (AAH) model~\cite{aubry1980,harper1955,jitomirskaya2015}, which becomes self-dual for a critical quasiperiodic potential strength at which the system undergoes a metal--insulator transition. 
At the self-dual point, the AAH chain hosts multifractal eigenstates~\cite{hiramoto1989,piéchon1995} and a singular continuous (Cantor-set) energy spectrum~\cite{last1994}, realizing the one-dimensional descendant of two-dimensional Hofstadter physics via dimensional reduction~\cite{hofstadter1976}.
From a theoretical perspective, the AAH critical point is a rare example of an exactly solvable, non-perturbative quantum critical phase. 
At the same time, 
generic perturbations, including anisotropy~\cite{yahyavi2019generalized,alluf2025duality}, longer-range hopping~\cite{roy2021fraction,biddle2011localization,deng2019one}, or commensurate superlattice potentials~\cite{padhan2022emergence,yi2025unveiling}, break this duality, destabilizing the critical behavior. This raises a fundamental question that remains largely unexplored: \emph{is the AAH criticality intrinsically fragile, or can it re-emerge in a controlled and universal manner under strong self duality-breaking perturbations?}

This question is particularly timely given recent experimental advances. Cold-atom platforms have realized AAH and Hofstadter Hamiltonians using optical superlattices and laser-assisted tunneling~\cite{aidelsburger2013,miyake2013}, while photonic waveguide arrays have directly observed localization transitions in quasiperiodic lattices~\cite{lahini2009,kraus2012}. Synthetic dimensions in cold atoms, photonics, and circuit-based systems further enable precise control over unit-cell structure, sublattice potentials, and hopping processes~\cite{boada2012}. These capabilities naturally motivate the study of periodic or superlattice modulations in quasiperiodic systems. Such modulations are straightforward to implement experimentally, yet their impact on critical quasiperiodic phases remains poorly understood. In particular, strong superlattice potential is often assumed to trivially gap or localize the system, obscuring the possibility of more subtle reorganizations of critical spectral weight.

In this Letter, we show that when additional onsite periodic modulation or superlattice potential with periodicity $N$ is applied to the \emph{critical} AAH chain, it leads to a striking and counterintuitive outcome. Although weak and intermediate potential immediately breaks self-duality of the model leading to the appearance of delocalized and localized states, in the strong potential limit self-duality can be restored, generating an \emph{emergent universal critical regime}.
We derive the self-duality condition within an emergent critical band. The fate of self-duality in this regime depends sensitively on the period $N$: for $N=2$, self-duality is fully restored across all bands in the strong-potential limit, whereas for $N\ge3$ it is generically only partial. 

We further show that this limitation for the latter cases can be overcome using Hamiltonian engineering. By introducing suitably engineered sublattice-dependent hopping terms, full self-duality can be restored simultaneously in all bands for arbitrary $N$. Analytical arguments and numerical results demonstrate that the resulting strong-modulation critical phase is continuously connected to the original AAH critical point and belongs to the same universality class, exhibiting multifractal eigenstates and singular continuous spectra, characterized by the Hausdorff dimension. 

\emph{Model.--}
We consider the Aubry-Andr\'e–Harper (AAH) chain with an additional $N$-periodic onsite potential. The Hamiltonian is given by \begin{eqnarray}
H=H_0+H_1,
\label{ham}
\end{eqnarray}
where $H_1 = \sum\limits_{i} (t c_{i+1}^\dagger c_{i} + h.c.) + \lambda\cos(2\pi \beta i + \phi) c_{i}^\dagger c_{i}$ is the AAH model and $H_0 = \sum\limits_{i} V_i c_{i}^\dagger c_{i}$ is the overall periodic modulation.
Here $t$ is the nearest-neighbor hopping amplitude, $\lambda$ denotes the quasiperiodic potential strength with irrational frequency $\beta$, and $\phi$ is a global phase. The onsite potential $V_i$ is a generic $N$-periodic onsite modulation with strengths forming a non-degenerate sequence $\{V_1,V_2,..,V_N\}$ within a period of $N$ sites which is repeated throughout the lattice.
The AAH model ($H_1$) shows a delocalization-localization phase transition at $\lambda=2t$ where the model is self-dual and at this point all single particle eigenstates become critical (multifractal)~\cite{aubry1980}. 
Additionally, the energy spectra as a function of $\beta \in [0,1]$ shows the celebrated Hofstadter butterfly (HB) structure~\cite{hofstadter1976}.
With an additional superlattice modulation immediately breaks the self-duality leading to the  disappearance of critical states and symmetric HB. However, we will show that the AAH criticality re-emerges in the limit of strong superlattice modulations which also leads to the multiplication of HBs within a band. For the numerical results presented here, we have chosen PBC and the system size $L=F_n$, the $n\textsuperscript{th}$ term in the Fibonacci sequence. The quasiperiodicty is achieved by assuming $\beta=F_{n-1}/F_n$ unless otherwise mentioned. 
As $n \to \infty$, $\beta$ approaches the golden mean.

\emph{Emergent criticality under $N$-periodic onsite modulation.--}
We consider the Hamiltonian shown in Eq.~\ref{ham} in the limit of strong periodic potential such that $|V_\ell - V_{\ell'}| \gg t,\lambda$ where $\ell,\ell'$ are two elements in the sequence $\{V_\ell \}$ in the $N$-periodic onsite potential as discussed previously. At this limit, $H_1$ can be treated as perturbation to $H_0$. The $N$-periodic potential introduces $N$ sublattices and we focus on a fixed $\ell_0\textsuperscript{th}$ sublattice sector to define projectors $P = \sum\limits_{m=0}^{L/N-1} \ket{Nm+\ell_0}\bra{Nm+\ell_0}$ with energy $E_P = V_{l_0}$ and $Q = 1 - P$. Hence, one can apply Schrieffer-Wolff (SW) transformation that gives an effective Hamiltonian in the $P$-projected subspace, given by, $H_{eff} = PH_0P + PH_1P + P H_1 Q
\frac{1}{E_P - H_0}
Q H_1 P + P H_1 Q
\frac{1}{E_P - H_0}
Q H_1 Q\frac{1}{E_P - H_0}Q H_1 P + .....$ up to several order in $H_1$. The leading order non-vanishing hopping contribution appears at $N\textsuperscript{th}$ order term in $H_1$ which implies that the effective hopping 
$t_{\text{eff}}^{l_0}=
\frac{t^{N}}{\prod_{r=1}^{N-1} (V_{\ell_0} - V_{\ell_0+r})}$
Hence, an effective Hamiltonian for the $\ell_0\textsuperscript{th}$ sector is given by (for details see Ref.~\cite{suppl}),
\begin{eqnarray}
H_{\text{eff}}^{{\ell_0}}
=
t_{\text{eff}}^{\ell_0}
\sum_m
\left[
p_{m+1}^\dagger p_m + \text{h.c.}
\right]
+ 
\lambda
\cos(2\pi \beta' m + \phi)\,
p_m^\dagger p_m,
\end{eqnarray}
where $p_m$'s are operators on the projected sector and $\beta'=N\beta$.
This is the emergent AAH model with renormalized hopping and quasiperiodicity. The self-duality condition gives the critical point at the $l_0\textsuperscript{th}$ band which is given by,
\begin{eqnarray}
\lambda_{c,\ell_0} = 2 |t_{\text{eff}}^{\ell_0}| = \frac{2{|t|}^{N}}{\prod_{r=1}^{N-1} |V_{\ell_0} - V_{\ell_0+r}|}.
\label{crit_gen}
\end{eqnarray}
In the following, we discuss two exemplary cases, namely, the 2-periodic and 3-periodic modulations to validate the emergent criticality explained here. 
\begin{figure}
    \centering
    \includegraphics[width=1.0\linewidth]{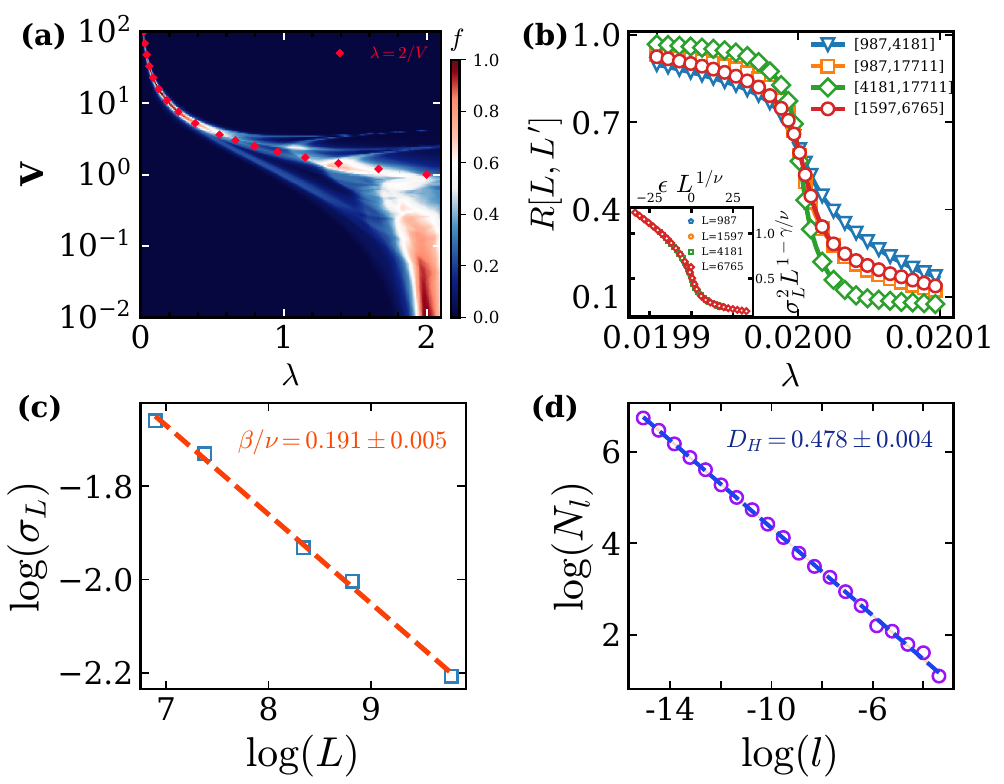}
    \caption{(a) Emergent criticality is shown as a function of quasiperiodic potential strength $\lambda$ and 2-periodic potential strength $V$ in terms of fraction of critical states ($f$) (shown in the color bar). AAH type criticality emerges at $\lambda\approx 2/V$ for large values of $V$ (red  diamonds).
    (b) $R[L,L']$ as a function of $\lambda$ for different system size-combinations $[L,L']$. The abscissa and ordinate of the crossing point gives the critical $\lambda_c\approx0.02$ and ratio $\gamma/ \nu\approx0.620 \pm 0.005$, respectively. The inset shows data collapse using finite-size scaling analysis for $\lambda_c$ and $\gamma/\nu$ obtained from figure (a). 
(c) $\log(\sigma_L)$ vs $\log(L)$ at $\lambda_c$ gives $\beta/\nu= 0.191 \pm 0.005$ from the slope of the fitted straight line.
(d) Hausdorff dimension $D_H$ is calculated from the plot of $\log(N_l)$ vs $\log(1/l)$. The slope from the fitted line (blue dashed) gives $D_H = 0.478 \pm 0.004$. }
\label{universality}
\end{figure}

\emph{Emergent criticality with 2-periodic modulations.--}
Here we consider the case of 2-period modulation, defined as a sequence $(V,~2V)$ for two consecutive sites repeated in a periodic manner. This sequence is chosen for convenience, since the theory works for any generic sequence. In the limit of $V\gg t, \lambda$ an effective Hamiltonian via the SW transformation up to second order in $H_1$ for the upper band (same as lower band) is, thus, given by~\cite{suppl},
\begin{eqnarray}
    H_{2p}
=
\frac{t^2}{V}
\sum_m
[
u_{m+1}^\dagger u_m + \text{h.c.}]
+
\lambda
\cos(4\pi \beta m + \phi)\,
u_m^\dagger u_m,
\label{2p,u}
\end{eqnarray}
which is the AAH model written in terms of operators $u_m$'s in the upper band subspace with doubled quasiperiodicity $\beta'= 2\beta$ and modified hopping $t'=\frac{t^2}{V}$. Eq.~\ref{2p,u} satisfies the self-duality condition $\lambda = 2t^2/V$ (assume $t=1$ in numerics) where the effective AAH model should become critical with the energy gaps exhibiting an HB structure. The presence of the periodic modulation immediately breaks the self-duality of the standard AAH model leading to appearance of delocalized and localized states in the energy spectrum. The self-duality gets restored in the limit of strong periodic potential $V$. To quantify the evolution of critical states in the parameter space, we resort to the inverse participation ratio which is defined, for a normalized single-particle eigenstate $\ket{\psi}$, as $\rm IPR=\sum\limits_i |\psi_i|^4\sim L^{-D_2}$, where the fractal dimension $D_2$ assumes values $1,~0$ and fraction for the delocalized, localized and critical states, respectively. Thus, we determine the fraction $f=N_c/N_t$ of critical states, where $N_c$ and $N_t$ are the numbers of critical states and all states, respectively, in the full spectrum (or within a band). We calculate $f$ over the full spectrum for different points in the $\lambda-V$ plane, which is shown in Fig.~\ref{universality}(a). It clearly shows how the AAH criticality, which breaks down for small and intermediate values of $V$, re-emerges in the strong limit of $V$ as one goes along the $\lambda=2/V$ line. This is also consistent with our analytical result (shown by the dashed line).  

\begin{figure}
    \centering
    \includegraphics[width=1.0\linewidth]{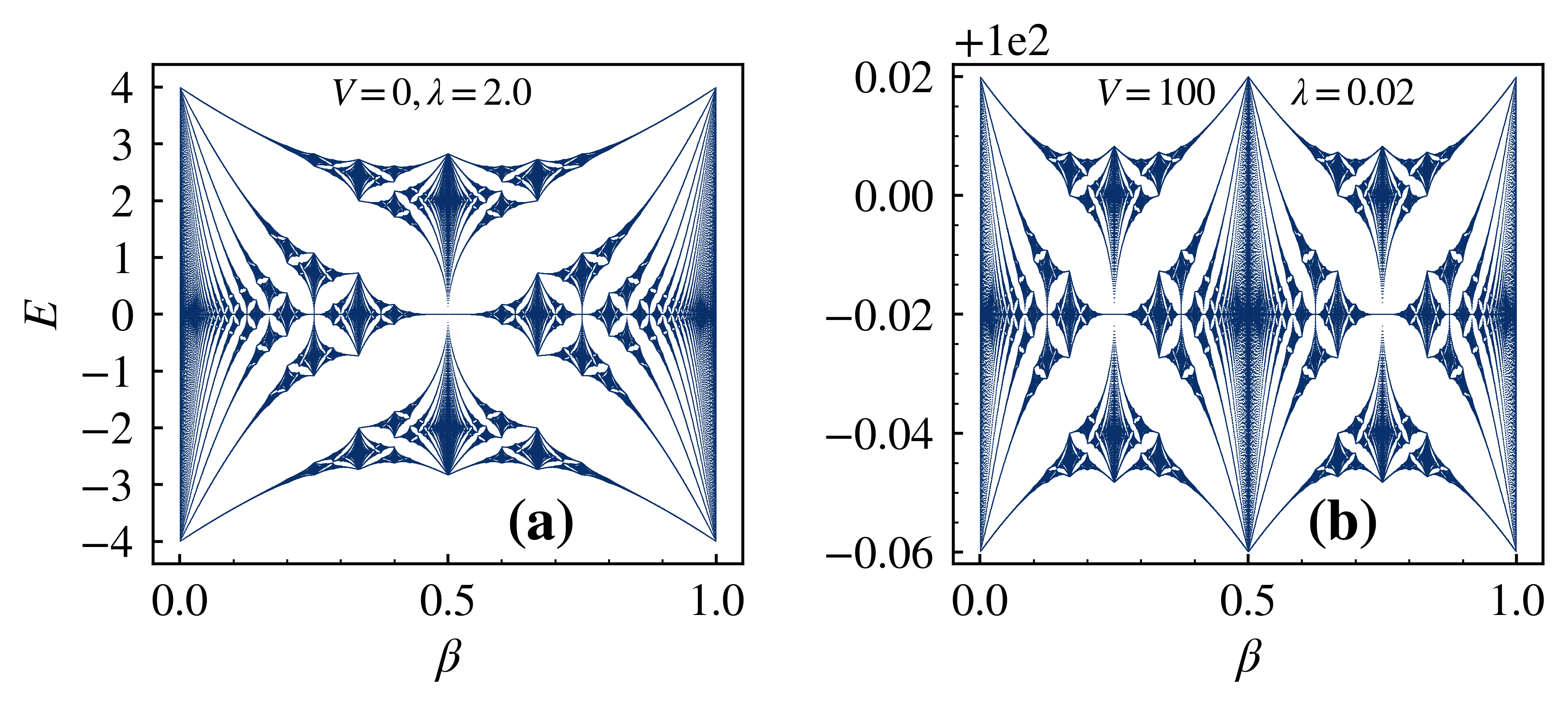}
    \caption{(a) The standard HB (energy spectrum as a function of $\beta$) at $V=0$ and $\lambda=2$, the self-dual AAH critical point.(b) HB doubling (shown for only the upper band) for 2-period superlattice modulation at large $V=100$ and $\lambda=0.02$ manifesting the emergent criticality at $\lambda=2/V$. For all plots, $L=2584$ and $\beta=\{1,2,...,L-1\}/L$.}
    \label{fig2}
\end{figure}

\emph{Critical analysis of emergent criticality.--}
In order to examine this emergent criticality in the system in the limit of strong $V$, we perform critical analysis of the states for 2-period modulation following a finite-size scaling method put forward in Refs.~\cite{hashimoto1992finite,roy2022critical}. 

Here we compute the order parameter for localization transition~\cite{roy2022critical} defined as $\sigma=\sqrt{\frac{\langle \rm{PR}\rangle}{L}}$, where  $\rm PR=1/\rm IPR$ is the participation ratio $\langle\cdot\rangle$ denotes the spectral average.
In the vicinity of the critical point $\lambda_c$ one defines the reduced parameter $\epsilon=|\lambda - \lambda_c|/\lambda_c$ such that the observables $\sigma\sim \epsilon^\beta$, $\rm{PR}\sim \epsilon^\gamma$ and the correlation length $\xi\sim \epsilon^\nu$ follow  a power-law scaling with the exponents $\beta$, $\gamma$ and $\nu$, respectively, that determine the universality class of the critical point $\lambda_c$. 

Following Ref.~\cite{roy2022critical}, we define a two-system size-variable function $R[L,L']=1 + [\ln(\sigma_L^2/\sigma_{L'}^2)/\ln(L/L')]$ and plot it as a function of $\lambda$ in the
vicinity of $\lambda_c$ for several pairs of the system
of sizes $L$ and $L'$ which intersect each other at a common fixed
point, as shown in Fig.~\ref{universality}(b) for a fixed $V=100$. The critical quasiperiodic potential strength $\lambda_c$ and the exponent ratio
$\gamma/\nu$ are determined from the abscissa and ordinate, respectively, of the
common crossing point. From Fig.~\ref{universality}(b) we obtain $\lambda_c=0.0199995$ $(\approx2/V)$ and $\gamma/\nu=0.620 \pm 0.003$. Assuming a finite-size scaling ansatz $\sigma^2=L^{\gamma/\nu - 1} G(\epsilon L^{1/\nu})$ (where $G$ is a scaling function), we are able to show, in the inset of Fig.~\ref{universality}(b), a data collapse near the critical point indicating that $\sigma^2 \sim L^{\gamma/\nu - 1} \epsilon L^{1/\nu}$, confirming $\nu=1$ and $\gamma=0.620 \pm 0.005$. Moreover, since $\xi=L$ at the critical point, it can be shown that $\sigma\sim L^{-\beta/\nu}$ at $\lambda_c$. From the straight line fit of $\log \sigma$ vs. $\log L$ plot at $\lambda_c$, as shown in Fig.~\ref{universality}(c), we obtain $\beta=0.191 \pm 0.005$. The exponents obtained in our analysis at the emergent criticality follow the hyperscaling relation $2\beta + \gamma = \nu$ of the phase transition, reflecting the consistency of our analysis. 

We also provide a characterization of the energy spectrum in the critical regime by computing the Hausdorff dimension of the system following a box-counting method~\cite{ikezawa1994energy}. Considering that the total number of boxes required is $N_l$ for a given box length $l$ such that $N_l$ spans over the entire energy spectrum, $N_l$ shows a power-law behavior with $l$ as $N_l\sim l^{-D_H}$, where $D_H$ denotes the Hausdorff dimension. Following this relation, we extract $D_H\approx0.5$ at the emergent critical point, here $V=100$ and $\lambda=0.02$, which is shown in Fig.~\ref{universality}(d). The extracted value is very close to $D_H=0.5$ obtained for standard criticality of the AAH model~\cite{ikezawa1994energy}.

Breaking of self-duality and its restoration is also reflected in the spectral topology which is shown in In Fig.~\ref{fig2}(a-b).  We show how the HB gap structure in standard AAH model i.e., at  $(\lambda=2,V=0)$ (Fig.~\ref{fig2}(a)), evolves into   two HBs in both bands at the emergent criticality $(\lambda=0.02,V=100)$ as shown in Fig.~\ref{fig2}(b) (for upper band only). This suggests that the self duality is restored in each band at the analytically predicted value $\lambda=2/V$ in the limit of strong periodic modulation.

\emph{Emergent criticality in bands with 3-period modulations.--}
Now, we discuss the case of 3-period modulation achieved by assuming potentials of strengths  $(V,~2V,~3V)$ at three consecutive sites repeated periodically. In the strong $V\gg t, \lambda$ limit, one obtains three almost flat bands far away from each other so that one can obtain three effective Hamiltonians corresponding to three bands via SW transformation up to third order in $H_1$. 
The effective Hamiltonian within a band is given by~\cite{suppl},  
\begin{eqnarray}
    H_{3p}
=
t_{\text{eff}}
\sum_m
[
p_{m+1}^\dagger p_m + \text{h.c.}] + 
\lambda
\cos(6\pi \beta m + \phi)\,
p_m^\dagger p_m,
\label{3p,band}
\end{eqnarray}
where, $p_m$'s are operators in the subspace of projected band~\cite{suppl}. This is again an emergent critical AAH model with triple quasiperiodicity $\beta_{eff}=3\beta$. Interestingly, however, the effective hopping in this case is band specific. We obtain that $t_{eff}=\frac{t^3}{2V^2}$ in both the upper and lower bands whereas $t_{eff}=-\frac{t^3}{V^2}$ in the middle band such that the self-duality and criticality is restored at $\lambda_c=2|t_{eff}|$. Hence, comparing with Eq.~\ref{crit_gen}, both the upper and lower bands become critical at $\lambda_{c_1}=\frac{t^3}{V^2}$ whereas the middle band shows criticality at a different point $\lambda_{c_2}=\frac{2t^3}{V^2}$ in the parameter space.

This peculiar behavior is shown in Fig.~\ref{fig3} (a) by plotting $D_2$ as a function of $\lambda$ and eigenstate index $n/L$ for $V=100$. The critical points $\lambda_{c_1}=1/100^2$ (for the upper and lower bands) $\lambda_{c_1}=2/100^2$ (for the middle band) are shown as the vertical dashed lines where $0<D_2<1$. The blue and maroon colored regions denote the delocalized and localised spectrum. This two critical points can also be seen by comparing $\langle \rm{IPR}\rangle$ and $\langle\rm{NPR}\rangle$ (where $\rm{NPR}=\rm{IPR}^{-1}/L$ is the normalized participation ratio) of the spectrum as a function of $\lambda$ as shown in Fig.~\ref{fig3} (b). Clearly when $\lambda_{c_1} < \lambda < \lambda_{c_2}$ (grey middle region) both $\langle \rm{IPR}\rangle$ (blue squares) and $\langle\rm{NPR}\rangle$ (orange circles) are finite. This complements the above fact that the localization transition does not occur at a single critical point unlike the previously discussed case for 2-period modulation. In Fig.~\ref{fig3}(c), we show how criticality emerges in the large $V$ limit by considering the middle band as an example. We plot the fraction of critical states $f$ defined earlier in the $\lambda$ - $V$ plane which shows the initial $\lambda=2$ at $V=0$ critical point gradually vanishes and reappears at higher values of $V$ which is in agreement with the emergent critical line $\lambda=2/V^2$ (with $t=1$), denoted as the dashed curve. We also provide a complementary analysis of the Hausdorff dimension $D_H$ which is $\approx 0.47$ (close to $0.5$) in the emergent critical midband, shown in Fig.~\ref{fig3}(d) as violet squares. We also show the Hausdorff dimension for the upper band at $\lambda_{c_2}$ (blue circles) which comes out as $D_H\approx 0.9$ (close to $1$), confirming its off-criticality. In this case also we obtain evolution of the critical state leading to splitting up of the HB ~\cite{suppl}.

In the following we will show that the criticality in the off-critical bands obtained for 3-period modulation case can be restored by appropriate Hamiltonian engineering. 

\begin{figure}
    \centering
    \includegraphics[width=1.0\linewidth]{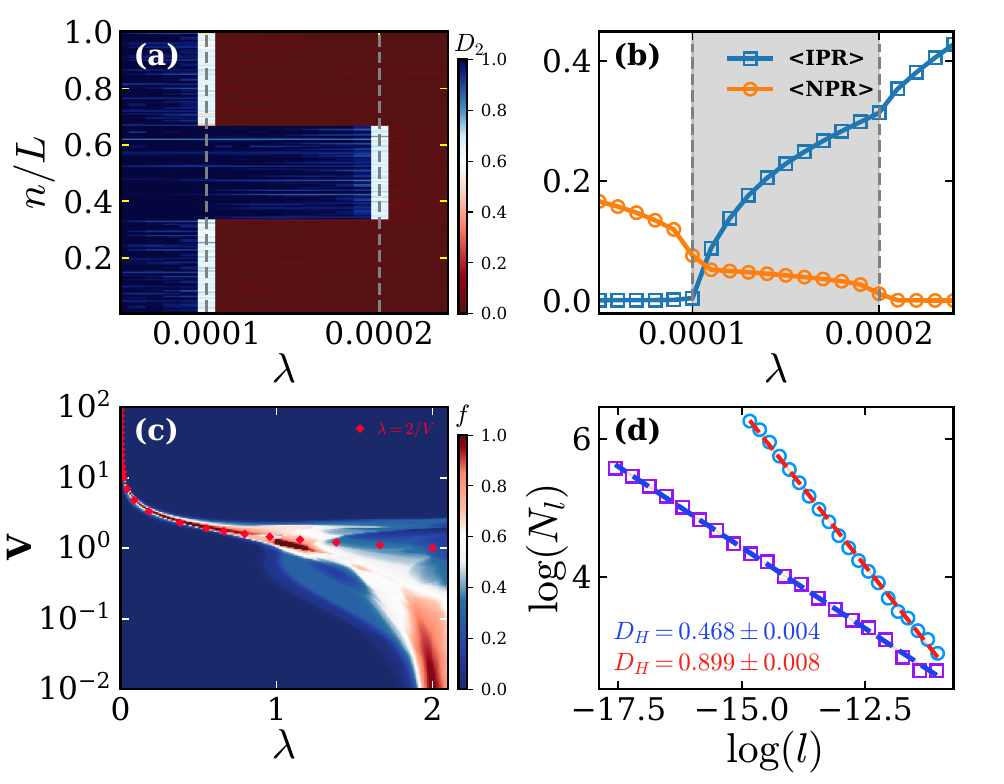}
    \caption{(a) The fractal dimension $D_2$ as a function of $\lambda$ and eigenstate index ($n/L$) at $V=100$ for 3-period superlattice modulation.  $\lambda_{c1}=1/V^2$ (left vertical line) and $\lambda_{c2}=2/V^2$ (right vertical line) are the band-dependent critical points (b) Spectrum-averaged $\langle \rm{NPR}\rangle$ (circles) and $\langle \rm{IPR}\rangle$ (squares) are plotted against $\lambda$, corresponding to (a), to demonstrate the presence of an intermediate phase between $\lambda_{c1}$ and $\lambda_{c2}$ (the gray-shaded region). (c) Emergence of criticality is shown as a function of quasiperiodic potential strength $\lambda$ and 3-periodic potential strength $V$ in terms of fraction of critical states $f$ for the middle band with $L=6765$. AAH criticality emerges at $\lambda\approx 2/V^2$ for large values of $V$ (red diamonds). (d) Hausdorff dimension $D_H=0.468 \pm 0.004$ (fitted blue dashed line) for the emergent critical middle band at $V=100$ and $\lambda_{c_2}=0.0002$. For the localized upper band, $D_H=0.899 \pm 0.008$ (fitted red dashed line). Data points are shown as violet squares (blue circles) for the middle (upper) band.}
    \label{fig3}
\end{figure}

\emph{Hamiltonian engineering for tuning criticality.--}
It is clear from Eq.~\ref{crit_gen} that different bands may have different critical points mainly due to different values of $t_{\text{eff}}^{\ell_0}$, which represents nearest neighbor hopping in $l_0\textsuperscript{th}$ sublattice giving rise to $l_0\textsuperscript{th}$ band. Hence, one can introduce intra-sublattice hoppings such that all bands will now have the same value for the total effective hopping strength $t_{\text{eff}}=t^\star$, say, which leads to the same critical point $\lambda_c=2|t^\star|$ for all the bands, in the strong periodic modulation limit. 
To be more precise, we can add hopping between the nearest neighbour (NN) identical sublattice sites as \(t_N^{(l)}\) with sublattice-index \(l=0,1,\dots,N-1\) (\(N\)-range hoppings). With the prescription, the strengths of these engineered sublattice hoppings are determined by:
$\; t_N^{(l)} \;=\; t^\star \;-\; t_{\rm eff}^{(l)} \;.\;$,
where $t_{\rm eff}^{(l)}$ is defined earlier. 
The engineered Hamiltonian is, thus, given by 
$H_{eng} = H + \sum\limits_{i=0}^{L-1} \bigl(t_N^{({l})}\, c_{i+N}^\dagger c_{i} + {\rm h.c.} \bigr)$,
where $l\equiv\bmod(i,N)$ and $H$ is given in Eq.~\ref{ham}.

In the example of 3-period modulation, as discussed previously, there are three sublattices with effective hoppings $t_{\rm eff}^{(0)}=\frac{t^3}{2V^2}$, $t_{\rm eff}^{(1)}=-\frac{t^3}{V^2}$ and $t_{\rm eff}^{(2)}=\frac{t^3}{2V^2}$ corresponding to upper, middle and lower bands, respectively, at the strong $V$ limit. Hence, adding NN sublattice hoppings with strengths $t_{3}^{(0)}=0$, $t_{3}^{(1)}=\frac{t^3}{2V^2}$ and $t_{3}^{(2)}=0$
in three sublattices would make total effective hopping identical for all bands, i.e. $t^\star=\frac{t^3}{2V^2}$. This implies  
emergent criticality via self-duality restoration  at $\lambda_c=\frac{t^3}{V^2}$ for all the bands. This is demonstrated in Fig.~\ref{fig4}(a) where we plot the fractal dimension $D_2$ of all eigenstates (where $n/L$ is the fractional eigenstate index) as a function of $\lambda$, after Hamiltonian-engineering. It can be seen that, after Hamiltonian engineering, the second critical point $\lambda_{c_2}$ disappears and the entire spectrum exhibits a sharp transition at $\lambda_{c_1}$ (compare with Fig.~\ref{fig3}(a). Similar behavior can also be seen by comparing $\langle \rm{IPR}\rangle$ and $\langle \rm{NPR} \rangle$ in Fig.~\ref{fig4}(b).  

Alternatively, we could have chosen the combination: $t_{3}^{(0)}=\frac{t^3}{2V^2}$, $t_{3}^{(1)}=0$ and $t_{3}^{(2)}=\frac{t^3}{2V^2}$ of added sublattice hoppings to target $t^\star=\frac{t^3}{V^2}$ such that $\lambda_c=\frac{2t^3}{V^2}$ for all bands, which is a different point in parameter space (see ~\cite{suppl} for detailed discussion). Hence, through the engineering of sublattice hoppings one can tune the emergent critical points which applies to even $(N>3)$-periodic onsite potential modulation of any generic form~\cite{suppl}.

\begin{figure}
    \centering
    \includegraphics[width=1.0\linewidth]{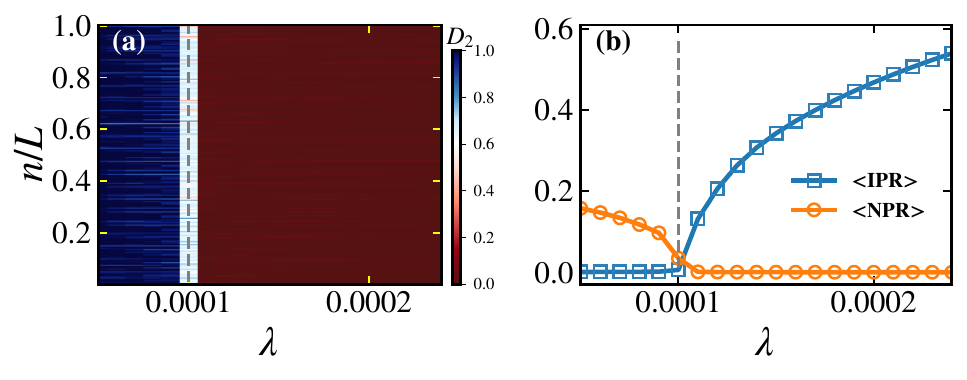}
    \caption{(a) $D_2$ after Hamiltonian-engineering for the $\lambda_{c1}$ where all eigenstates become critical. (c) Spectrum-averaged $\langle \rm{NPR}\rangle$ (circles) and $\langle \rm{IPR}\rangle$ (squares) are plotted against $\lambda$, corresponding to figure (a), which shows the shrinking of the intermediate phase to a single critical point at $\lambda_{c1}$. }
    \label{fig4}
\end{figure}

\emph{Conclusion.--}
We have shown that applying the $N$-period onsite modulations to the critical Aubry-Andr\'e-Harper chain first destroys the criticality and then generates an emergent universal critical regime in the limit of strong modulating onsite potential. We proved that the vanishing and revival of the criticality is associated with the break-down and subsequent restoration of the self-duality of the AAH model with increase in the onsite potential modulation. We characterized such emergent critically through  multifractal eigenstates, Hofstadter butterfly multiplication, and singular continuous spectra (Hausdorff dimension). For $N=2$, we have shown that the self-duality is fully restored across all bands, while for $N\ge3$, full self-duality can be engineered through sublattice-dependent hopping, demonstrating a general mechanism for Hamiltonian-engineered control of criticality. This emergent criticality is continuously connected to the original Harper universality class, revealing that quasiperiodic critical phases can survive and reorganize under strong self duality-breaking perturbations. Our predictions are directly accessible in already existing solid state, cold-atom, photonic, and electric circuit platforms~\cite{roati2008,modugno2009,aidelsburger2013,miyake2013,lahini2009,kraus2012,celi2014,an2017,stuhl2015}. These findings open up opportunities to explore engineered multifractal matter, controllable nonergodic dynamics, band-selective transport, and criticality-based parameter sensing. Thus, our work establishes the quasiperiodic systems with periodic modulations or superlattice potential as versatile platforms for both fundamental studies and quantum simulation applications.

\emph{Acknowledgments.-}T.M. acknowledges support from Science and Engineering Research Board (SERB), Govt. of India, through project No. MTR/2022/000382 and STR/2022/000023.

\bibliography{refs}

\clearpage
\newpage

\widetext

\appendix
\renewcommand{\appendixname}{}
\renewcommand{\thesection}{{S\arabic{section}}}
\renewcommand{\theequation}{S\arabic{equation}}
\renewcommand{\thefigure}{S\arabic{figure}}

\setcounter{section}{1}
\setcounter{figure}{0}
\setcounter{equation}{0}
\onecolumngrid

\begin{center}
\textbf{Supplementary materials for ``Emergent criticality in the Aubry-Andr\'e model with periodic modulation"}
\end{center}

\vspace{5mm}

\author{Sitaram Maity}
\thanks{These authors contributed equally}
\author{Nilanjan Roy}
\thanks{These authors contributed equally}
\author{Tapan Mishra}
\email{mishratapan@niser.ac.in}

\affiliation{School of Physical Sciences, National Institute of Science Education and Research, Jatni,  Odisha 752050, India}
\affiliation{Homi Bhabha National Institute, Training School Complex, Anushaktinagar, Mumbai, Maharashtra 400094, India}

\date{\today}

\maketitle

\section{\thesection: Derivation of the Schrieffer-Wolff Hamiltonian for Generic $q$-Period Modulation}
We consider a one-dimensional Aubry-Andr\'e-Harper (AAH) chain with a generic $q$-period onsite modulation. The $q$-period sequence $(V_0,V_2,...,V_{q-1})$ can also be written in terms of site-index $n$ as:
\begin{equation}
V_n = V_{\ell}, \qquad \ell \equiv n \bmod q ,
\end{equation}
where the onsite energies $V_\ell$ are all distinct and satisfy
$|V_\ell - V_{\ell'}| \gg t,\lambda$.
The Hamiltonian is given as 
\begin{equation}
H = H_0 + H_1 ,
\end{equation}
with
$H_0 = \sum_n V_{n}\, c_n^\dagger c_n$,
and
$H_1=t \sum_n\left(c_{n+1}^\dagger c_n + \text{h.c.}\right)+
\lambda \sum_n
\cos(2\pi \beta n + \phi)\,
c_n^\dagger c_n$.
Due to the construction of the Hamiltonian, the Hilbert space decomposes into $q$ sublattices:
\begin{equation}
\mathcal H = \bigoplus_{\ell=0}^{q-1} \mathcal H_\ell ,
\qquad
\mathcal H_\ell = \{\, \ket{qm+\ell} \text{with }  m=0,1,2,..,L/q-1 \,\}.
\end{equation}
Each subspace has unperturbed energy
$E_\ell = V_\ell$.
We focus on a fixed sector $\ell_0$ and define projectors
\begin{equation}
P = \sum_m \ket{qm+\ell_0}\bra{qm+\ell_0},
\qquad
Q = 1 - P .
\end{equation}
In the following we derive terms of different order contribution to the Schrieffer-Wolff (SW) Hamiltonian.

\subsection*{Zeroth Order}

The zeroth-order effective Hamiltonian is
\begin{equation}
H_{\text{eff}}^{(0)} = P H_0 P
= E_{\ell_0} \sum_m d_m^\dagger d_m ,
\end{equation}
where
$d_m = c_{qm+\ell_0}$. This is constant term and hence, can be ignored. 

\subsection*{First Order}

Nearest-neighbor hopping changes the sublattice index by $\pm 1 \bmod q$, so projecting the quasiperiodic potential onto the $\ell_0$ sublattice gives
\begin{equation}
H_{\text{eff}}^{(1)}=P H_1 P = \lambda \cos(2\pi \beta (qm+\ell_0) + \phi)
=
\lambda \cos(2\pi q \beta m + \phi + 2\pi \beta \ell_0) .
\end{equation}
Thus,
$\beta_{\text{eff}} = q\beta$.

\subsection*{Second Order}

The second-order Schrieffer--Wolff Hamiltonian is
\begin{equation}
H_{\text{eff}}^{(2)}
=
P H_1 Q
\frac{1}{E_{\ell_0} - H_0}
Q H_1 P .
\end{equation}

All two-step virtual processes begin and end on the same site
\begin{equation}
\ket{qm+\ell_0}
\rightarrow
\ket{qm+\ell_0 \pm 1}
\rightarrow
\ket{qm+\ell_0},
\end{equation}
and therefore produce only onsite energy renormalization.
Thus, no inter-cell hopping is generated at second order.

\subsection*{Leading Nontrivial Contribution}

At order $k$, Schrieffer--Wolff transformation generates virtual paths of length $k$
connecting the target subspace to itself.
Each nearest-neighbor hop changes the sublattice index by $\pm 1 \bmod q$.
Therefore, a nontrivial process connecting
\begin{equation}
\ket{qm+\ell_0} \rightarrow \ket{q(m+1)+\ell_0}
\end{equation}
must traverse all intermediate sublattices:
\begin{equation}
\ell_0
\rightarrow
\ell_0+1
\rightarrow
\ell_0+2
\rightarrow
\cdots
\rightarrow
\ell_0+q-1
\rightarrow
\ell_0 \quad (\text{mod } q).
\end{equation}

This requires $q-1$ virtual excursions into $Q$.\\

The first nonvanishing hopping contribution appears at order $q-1$:
\begin{equation}
\begin{aligned}
H_{\text{eff}}^{(q-1)}
=&
P H_1 Q
\frac{1}{E_{\ell_0} - H_0}
Q H_1 Q
\frac{1}{E_{\ell_0} - H_0}
\cdots
Q H_1 P .
\end{aligned}
\end{equation}

The corresponding virtual path is
\begin{equation}
\ket{qm+\ell_0}
\rightarrow
\ket{qm+\ell_0+1}
\rightarrow
\cdots
\rightarrow
\ket{qm+\ell_0+q-1}
\rightarrow
\ket{q(m+1)+\ell_0}.
\end{equation}

The effective hopping amplitude scales as
\begin{equation}
t_{\text{eff}}
=
\frac{t^{q}}{\prod_{r=1}^{q-1} (E_{\ell_0} - E_{\ell_0+r})}=\frac{t^{q}}{\prod_{r=1}^{q-1} (V_{\ell_0} - V_{\ell_0+r})}.
\label{crit_hop}
\end{equation}

\subsection*{Final Effective Hamiltonian}

Up to a constant energy shift, the effective Hamiltonian in $\ell_0$ sector is

\begin{eqnarray}
H_{\text{eff},\ell_0}
=
t_{\text{eff}}^{\ell_0}
\sum_m
\left(
d_{m+1}^\dagger d_m + \text{h.c.}
\right)
+
\lambda
\sum_m
\cos(2\pi q \beta m + \phi + 2\pi \beta \ell_0)\,
d_m^\dagger d_m.
\end{eqnarray}
This is the AAH model with renormalized hopping and quasiperiodicity $\beta_\text{eff}=q\beta$ and constant-shifted phase $\phi'=\phi + 2\pi \beta \ell_0$.
The self-duality condition gives the critical point at the $l_0\textsuperscript{th}$ band which is given by,
\begin{eqnarray}
\lambda_{c,\ell_0} = 2 |t_{\text{eff}}^{\ell_0}| = \frac{2|t|^{q}}{\prod_{r=1}^{q-1} |V_{\ell_0} - V_{\ell_0+r}|}.
\label{crit_gen}
\end{eqnarray}
In the main text we have demonstrated two exemplary cases: one with the 2-period modulation sequence $(V,~2V)$ and the other with the 3-period modulation sequence $(V,~2V,~3V)$, repeated on the lattice in a periodic manner. Our derivation is applicable for any non-degenerate sequence of generic form and the choices of sequence taken here are purely for convenience. In the following we will discuss the evolution of critical states and spectral topology in detail for these two cases.
\setcounter{section}{2}
\section{\thesection: Evolution of Critical States and Hofstadter Butterfly}
We compute the energy-resolved fractal dimension \(D_2\) of single-particle eigenstates for the one-dimensional Aubry–André–Harper (AAH) model with an added periodic onsite potential. The numerical procedure used to produce the heatmaps (Fig.~\ref{fig:critical_states_evolution_period-2}, panels (a–d)) is as follows.

\paragraph{Energy normalization and binning.}
For each system size the full single-particle spectrum is linearly rescaled to the unit interval \(n/L\in[0,1]\), where $n/L$ is the normalized eigenstate index with $L$ system size and $n$ ranging from 0 to $L-1$. This normalization makes spectra from different \(L\) directly comparable and avoids spurious shifts when combining data. The normalized spectrum is partitioned into \(N_{\mathrm{bins}}=200\) equal-width energy bins. For a given bin \(b\), we collect all eigenstates whose normalized energy falls into that interval.

\paragraph{Extracting \(D_2\) by finite-size scaling.}
We assume the usual scaling form for the second generalized dimension as mentioned in the main text, \(\langle\mathrm{IPR}\rangle_b \;\propto\; L^{-D_2(b)}\), so that:
\begin{equation}\label{eq:fit}
\log\langle\mathrm{IPR}\rangle_b \;=\; -D_2(b)\,\log L + \mathrm{const}.
\end{equation}
For each $b$ we estimate \(D_2\) as minus the slope of the linear fit of \(\log\langle\mathrm{IPR}\rangle_b\) versus \(\log L\). 

\paragraph{Hamiltonian and parameter sampling.}
For each pair \((\lambda,V)\) the single-particle Hamiltonian is diagonalized on lattices of sizes \(L\in\{610,\,2584,\,10946\}\). The three sizes are chosen to allow a simple finite-size scaling estimate of \(D_2\) according to Eq.~\ref{eq:fit}. The main text describes the Hamiltonian; here we use the same notation: \(\lambda\) denotes the amplitude of the incommensurate AAH potential and \(V\) denotes the amplitude of the alternating periodic onsite potential.

\paragraph{Evolution of Hofstadter Butterfly(HB).}
We show the evolution of HB of gap structure for various parameter sets $(\lambda, V)$ by calculating energy spectrum as a function of parameter $\beta \in [1,2,..,L-1]/L$ as mentioned in the main text.


The AAH self-duality enforces a global, energy-independent localization transition. Introducing a periodic sublattice potential (of strength $V$) explicitly breaks this microscopic symmetry; different eigenstates (which probe different spatial scales and sublattice structure) couple differently to the periodic potential. Intuitively, states whose envelope overlaps strongly with the periodic modulation see a different effective potential than states concentrated on the complementary sublattice, producing energy-dependent thresholds and thus mobility edges. As \(V\) becomes the dominant scale, the low-energy/long-wavelength effective model simplifies and an emergent duality governed by a dominant scale ($1/V$ or $1/V^2$) can restore a near-uniform criticality across the isolated bands. Thus accordingly with increasing $V$ the structure of the HB modifies and splits into q sub-bands.

\subsection{The 2-period modulation case ($V, 2V$)}

We first analyze the 2-period modulation. This configuration bisects the Brillouin zone, fundamentally restructuring the energy spectrum into two primary subbands separated by a gap at half-filling. Crucially, because the potential strictly alternates between $V$ and $2V$, the system preserves its chiral (particle-hole) symmetry around the band center.

\subsubsection{Weak to Intermediate Regime: The Emergence of the Mobility Edge}
As the potential strength is turned on from $V=0.0$ to $V=0.2$ and $V=2.0$ (Fig.~\ref{fig:critical_states_evolution_period-2}, panel a-c), the pristine AAH self-duality breaks down. The HB structure started to split from the middle of the spectrum, opening gaps at the centre of the band (Fig.~\ref{fig:critical_states_evolution_period-2}, panel e-g). Fig.~\ref{fig:critical_states_evolution_period-2}(panels b-c, f-g) visually captures this intermediate regime. The phase boundary between the extended states ($D_2 \approx 1$) and localized states ($D_2 \approx 0$) develops a highly complex, jagged structure leading to creation of mobility edges the number of which is larger for smaller $V$, as shown in Fig.~\ref{fig:critical_states_evolution_period-2}(a). 
Because of the preserved chiral symmetry, the mobility edges are perfectly symmetric around the central normalized state index $n/L = 0.5$. The pronounced ``fingers" of extended states that pierce into the localized regime (surviving up to nearly $\lambda = 2.0$) represent specific energy resonances where the destructive interference of the incommensurate disorder $\lambda$ is partially mitigated by the 2-period onsite potential. Presence of this symmetry splits butterfly spectra symmetrically around $V$ as we see in the 2nd rows of Fig.~\ref{fig:critical_states_evolution_period-2}. At $V=2.0$ and $\lambda=1.1$ in Fig.~\ref{fig:critical_states_evolution_period-2}(g), the (asymmetric) HB have finite smaller overlap near $\beta=0,0.5$ and $1.0$.

\subsubsection{Strong Potential Limit: Subband Isolation and $1/V$ Scaling}
\begin{figure}[t!]
    \centering
    \includegraphics[width=1.0\linewidth]{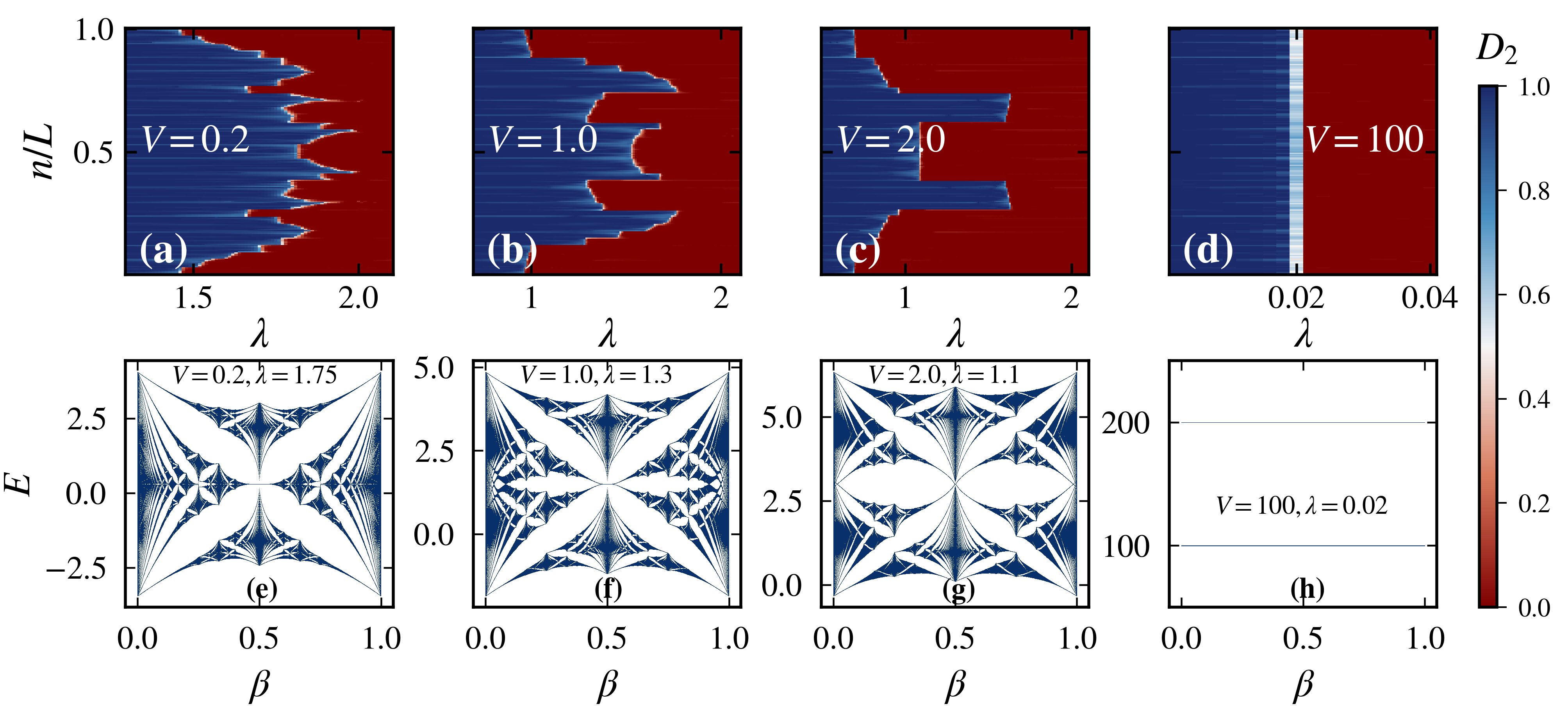}
    \caption{{\bf Evolution of Critical States and Hofstadter Butterfly with 2-period modulation:} 
    Energy-resolved fractal dimension $D_2$, shown for increasing values of $V$ from in figures (a-d). The vertical axis is the normalized eigenstate index $n/L\in[0,1]$; the horizontal axis is the AAH disorder strength $\lambda$. The system sizes used to estimate $D_2$ were $L=\{610,\,2584,\,10946\}$. Energy spectrum ($E$) as a function of $\beta$ is shown at various ($V,\lambda$) points in figures (e–h) mapping the transition from (e) close to standard AAH limit to (h) the same at the emergent AAH limit for  large $V$ for system size $L=2584$.}
    \label{fig:critical_states_evolution_period-2}
\end{figure}

As $V$ is pushed into the strong coupling regime, the energy gap at half-filling widens massively, and the two subbands become completely isolated. 
Within the strongly decoupled $V$ subband, an electron situated at site $i$ can no longer directly hop to site $i+1$ due to the massive energy penalty of the $2V$ site. Instead, transport occurs via a second-order virtual hopping process to site $i+2$. Using degenerate perturbation theory, the effective hopping parameter scales as $t_{\text{eff}} \propto t^2/V$, precisely $t_{\text{eff}}= t^2/V$ which is mentioned in the main text. Because the subband is physically isolated, the system behaves as a renormalized standard AAH model with hopping $t_{\text{eff}}$. The exact self-duality is dynamically restored within this subband, and the critical  point shifts to $\lambda_c = 2t_{\text{eff}}$. Therefore, the transition point strictly follows a first-order inverse power law: $\lambda_c = 2/V$ $(t=1)$, as depicted in Fig.~\ref{fig:critical_states_evolution_period-2}(d)) for $V=100$. At this point the two subbands become almost flat (see Fig.~\ref{fig:critical_states_evolution_period-2}(g))) with 2-period symmetric HB structure in each band, a zoomed-in version of which is shown in the main text for the upper band.

\subsection{The 3-period modulation case ($V, 2V, 3V$)}

\begin{figure}[t!]
    \centering
    \includegraphics[width=1.0\linewidth]{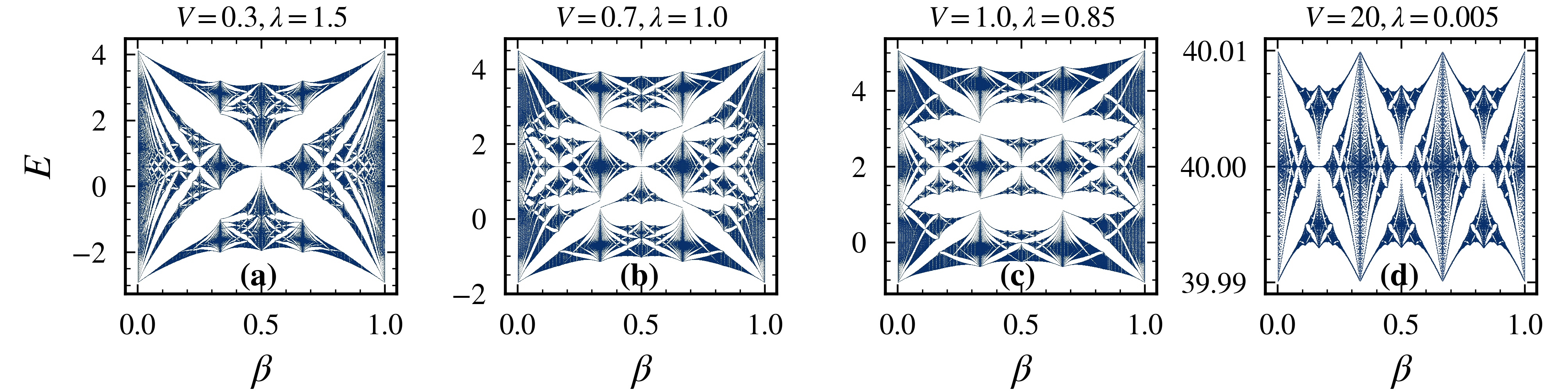}
    \caption{{\bf Evolution of Hofstadter Butterfly with 3-period modulation:} 
    Energy spectrum ($E$) as a function of $\beta$ is shown at various ($V,\lambda$) points across the phase (panels a-d). Each panel corresponds to a fixed ($V,\lambda$) mapping the transition from (a) close to the AAH limit to (d) the strictly isolated Middle Band at large $V$ for system size $L=987$.}
    \label{fig:critical_states_evolution_period-3}
\end{figure}

We now turn to the 3-period onsite potential. This geometry inherently breaks the bipartite symmetry of the lattice, dividing the spectrum into three distinct subbands centered near $V$, $2V$, and $3V$. In this case we have band-dependent criticality which is discussed in the main text. The evolution of critical states and mobility edges with a particular band (middle band, say) look very similar to the behavior described for 2-period modulation case and hence, is not, separately, discussed here. However, we will briefly discuss the evolution of the Hofstadter butterfly spectrum as a function of the quasi-periodicity $\beta$ (Fig.~\ref{fig:critical_states_evolution_period-3} panels a-d).

\subsubsection{Weak Regime: Spectral Deformation}
For weak potentials ($V=0.1$ to $V=0.3$), the trimerization acts as a perturbation that fragments the main energy band. While the 3-period potential breaks the standard bipartite symmetry of the pristine lattice, the specific balanced distribution of the onsite energies $\{V, 2V, 3V\}$ preserves a macroscopic $E \leftrightarrow -E$ spectral reflection symmetry around $E=2V$. 
As a consequence, the pristine, single-envelope Hofstadter butterfly undergoes significant deformation (Fig.~\ref{fig:critical_states_evolution_period-3}, panels a-d). Because the unperturbed 3-period potential naturally folds the Brillouin zone and opens gaps, the fractal spectrum begins to pinch symmetrically, showing incipient splitting and new local  gap-structures near $E=2V$ preparing the spectrum for full fracturing.

\subsubsection{Intermediate Regime: Subband Formation}


In the intermediate $V\sim 1$ regime, the Hofstadter spectrum visibly ruptures (Fig.~\ref{fig:critical_states_evolution_period-3}, (b–c)) into three distinct macroscopic lobes: an upper band centered around $E=3V$, a lower band around $E=V$, and a central band tightly bound around $E=2V$. Each macroscopic subband begins to host its own internal, miniature (asymmetric) Hofstadter-like fractal structure. At $V=1$ and $\lambda=0.85$ (Fig.~\ref{fig:critical_states_evolution_period-3},(c)), the three subbands are well separated except at $\beta=0$, and $1$.

\subsubsection{Strong Regime and Large-$V$ Limit: Decoupling and Hofstadter Butterfly Restoration}
The fundamental shift in physics occurs at strong $V$. 
At $V=20$ and $\lambda=2/V^2$, the main spectrum completely fractures into three independent subbands with self-duality being restored in the middle (critical) band whereas the other bands remain off-critical at this point. 
The 3-period (symmetric) Hofstadter Butterfly spectrum in the middle band  in Fig.~\ref{fig:critical_states_evolution_period-3}(d) which is  consistent with the emergent criticality condition derived analytically and the IPR-based results shown in the main text.


\end{document}